# Ultra-low energy (20 – 300 eV) electron imaging and diffraction

Lecture at WiSe 2026 School in Jülich

Tatiana Latychevskaia
tatiana@physik.uzh.ch
Paul Scherrer Insitute, Forschungstrasse 111, 5323 Villigen, Switzerland
Department of Physics, University of Zurich, Winterthurerstrasse 190, 8057 Zurich, Switzerland

## Contents

## 1. Properties of ultra low-energy electrons

### 1.1 Relatively large wavelength

Ultra-low electron energies range between 20 and 300 eV. The de Broglie electron wavelength is given by [1]:

$$\lambda = \frac{hc}{\sqrt{eU\left(2m_0c^2 + eU\right)}}, \tag{1}$$

where $h$ is the Planck's constant, $c$ is the speed of light, $m_0$ is the relativistic mass of the electron, $e$ is an elementary charge, and $U$ is the electron energy in eV. From Eq. (1), it can be seen that as the energy of the electron is decreasing, its wavelength is increasing. In conventional transmission electron microscopes (TEMs), the energy of electrons ranges between 80 and 300 keV, and thus, their wavelength 1.97 – 4.17 pm allows for imaging at atomic resolution. Low-energy electrons, on the other hand, have very large wavelength when compared to the electrons in commercial TEMs. For example, the wavelength of 150 eV electrons is one Ångstrom.

### 1.2 Short inelastic mean free path (IMFP)

Low-energy electrons interact with matter much stronger than high-energy electrons. The inelastic mean free path (IMFP) universal curve exhibits the minimum around 100 eV, which is exactly at the electron energy typical for low-energy electron imaging [2], Fig. 1. The transmission of low-energy electrons through a few layers of graphene showed that the intensity of the transmitted electrons decreased by about 25% with each additional graphene layer, with very weak intensity detected through three layers of graphene [3]. It means that only very thin samples can be imaged by low-energy electrons in transmission mode, which is maybe one of the disadvantages of ultra-low-energy electrons.

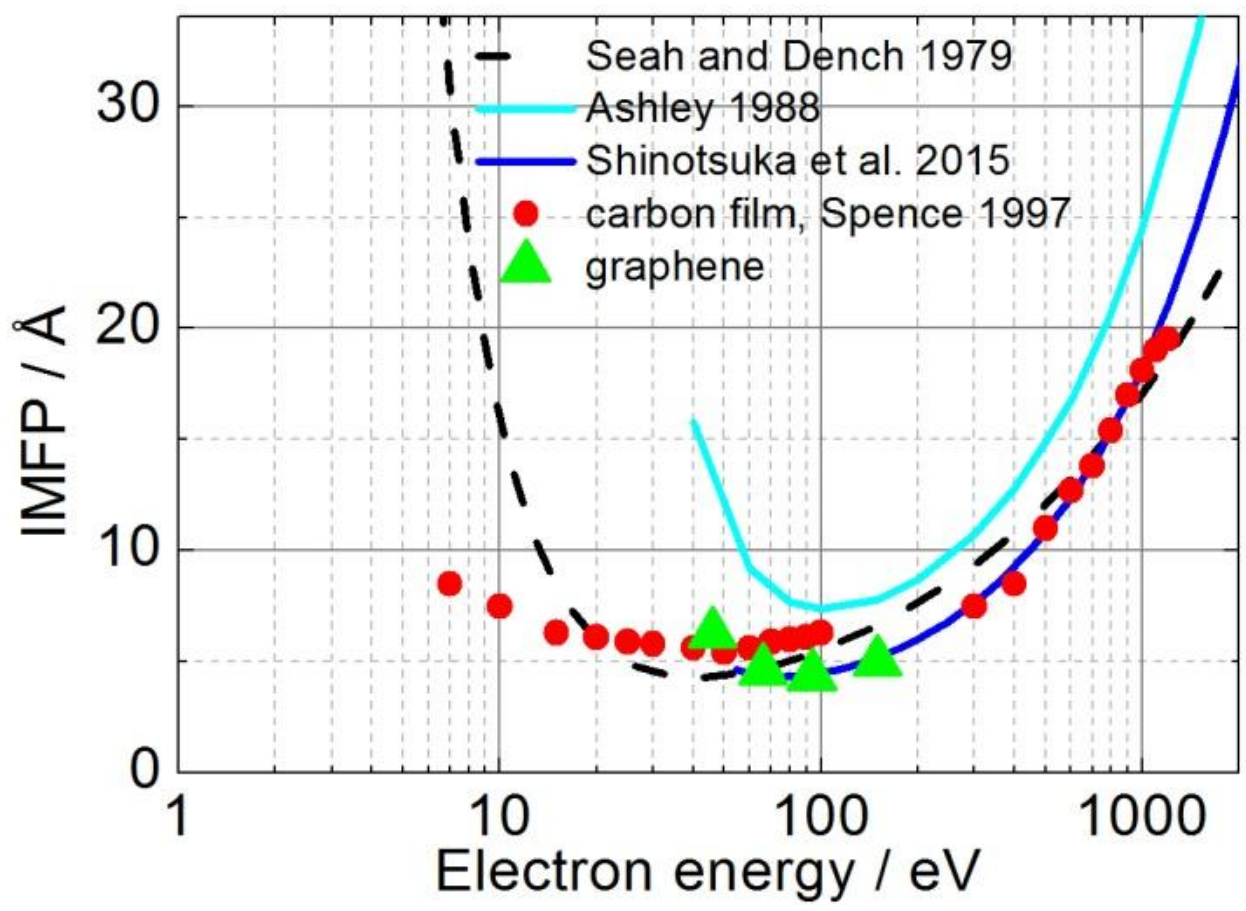

Fig. 1. Inelastic mean free path (IMFP) as a function of electron energy. (a) IMFP calculated by Seah and Dench [4], for carbon by Ashley [5], for graphite by Shinotsuka et al. [6], and experimentally measured as transmission through carbon films by Spence [7] and through graphene [2]. Adapted from [2].

## 1.3 Scattering amplitude and backward scattering

Another important difference between the electrons with ultra-low and conventional energies is the distribution of their scattering amplitudes. The scattered wave found by solving the Schrodinger equation in the first approximation is given by:

$$\psi(\vec{r}) = Ae^{ikz} + f(\vartheta,\varphi)A\frac{e^{ikr}}{r}, \tag{2}$$

where the incident wave is a plane wave of amplitude *A*, the scattered wave is a spherical wave with the complex-valued scattering amplitude $f(\vartheta,\varphi)$, and $\vec{r} = (x,y,z)$ is the coordinate. The distribution of the scattering amplitude $f(\vartheta,\varphi)$ depends on the energy and the chemical origin of the scattering atom. It can be expressed through a series:

$$f(\vartheta) = \sum_{l=0}^{\infty}(2l+1)f_l(k)P_l(\cos\vartheta), \tag{3}$$

where $P_l(\cos\vartheta)$ are the Legendre polynomials,

$$f_l(k) = \frac{1}{k}e^{i\delta_l(k)}\sin[\delta_l(k)], \tag{4}$$

and $\delta_l(k)$ are the phase shifts that can be found in the online NIST electron elastic scattering database [8]. The differential cross-section is given by:

$$\frac{\mathrm{d}\sigma}{\mathrm{d}\Omega}=\left|f\left(\vartheta\right)\right|^{2}. \tag{5}$$

The distribution of the differential cross-section as a function of the scattering angle $\vartheta$, exhibits a maximum in the forward direction (at $\vartheta = 0°$). In the case of ultra-low-energy electrons, the amplitude exhibits a second maximum – in the backward direction (at $\vartheta = 180°$), Fig. 2 and 3. This makes it possible to use low-energy electron diffraction in back-scattered mode, for example, as it is done in low-energy electron diffraction (LEED).

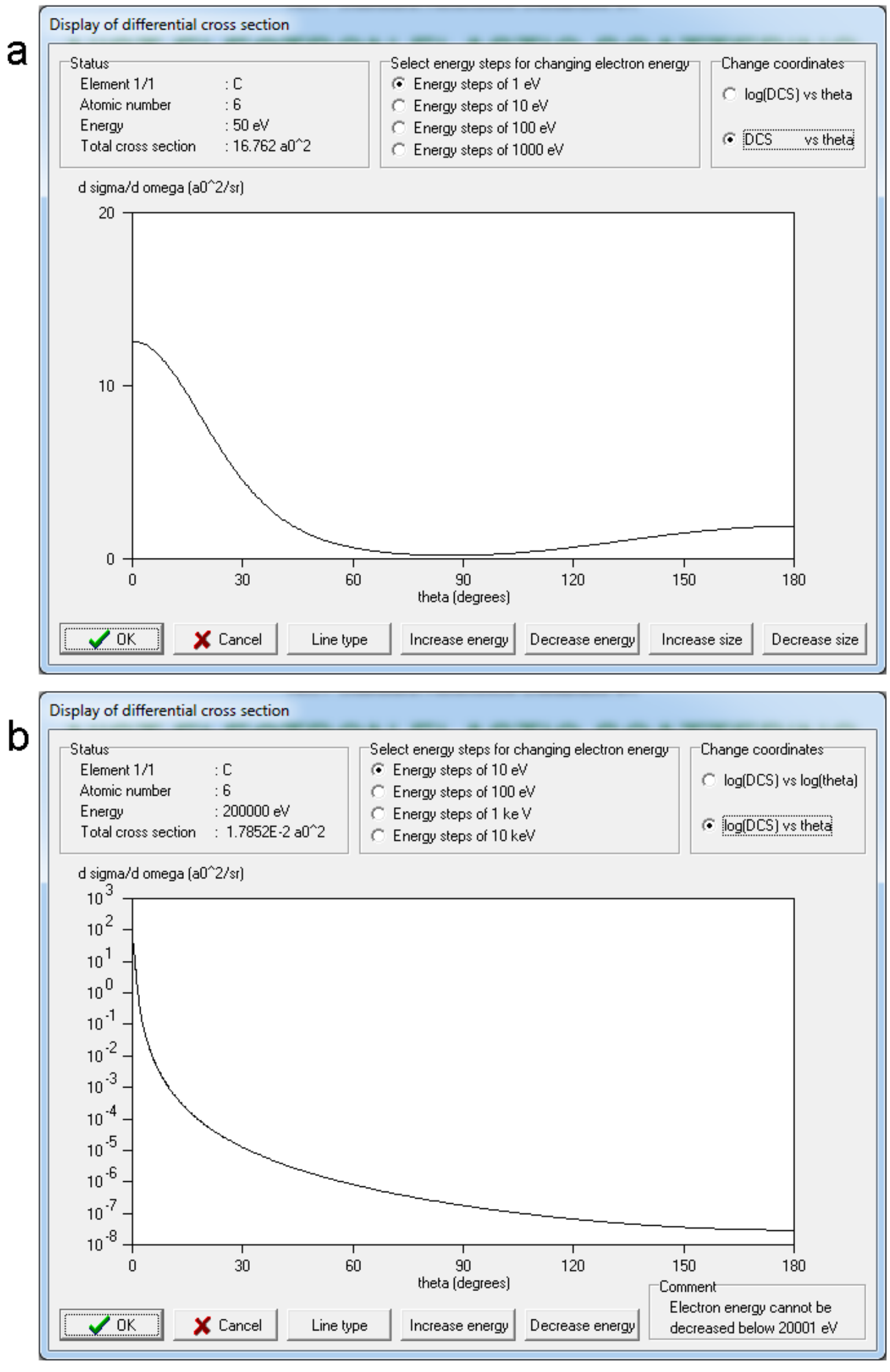

Fig. 2. Scattering differential cross-section (DCS) of electrons of (a) 50 eV and (b) 200 keV kinetic energy scattered by a carbon atom; calculated using the NIST database.

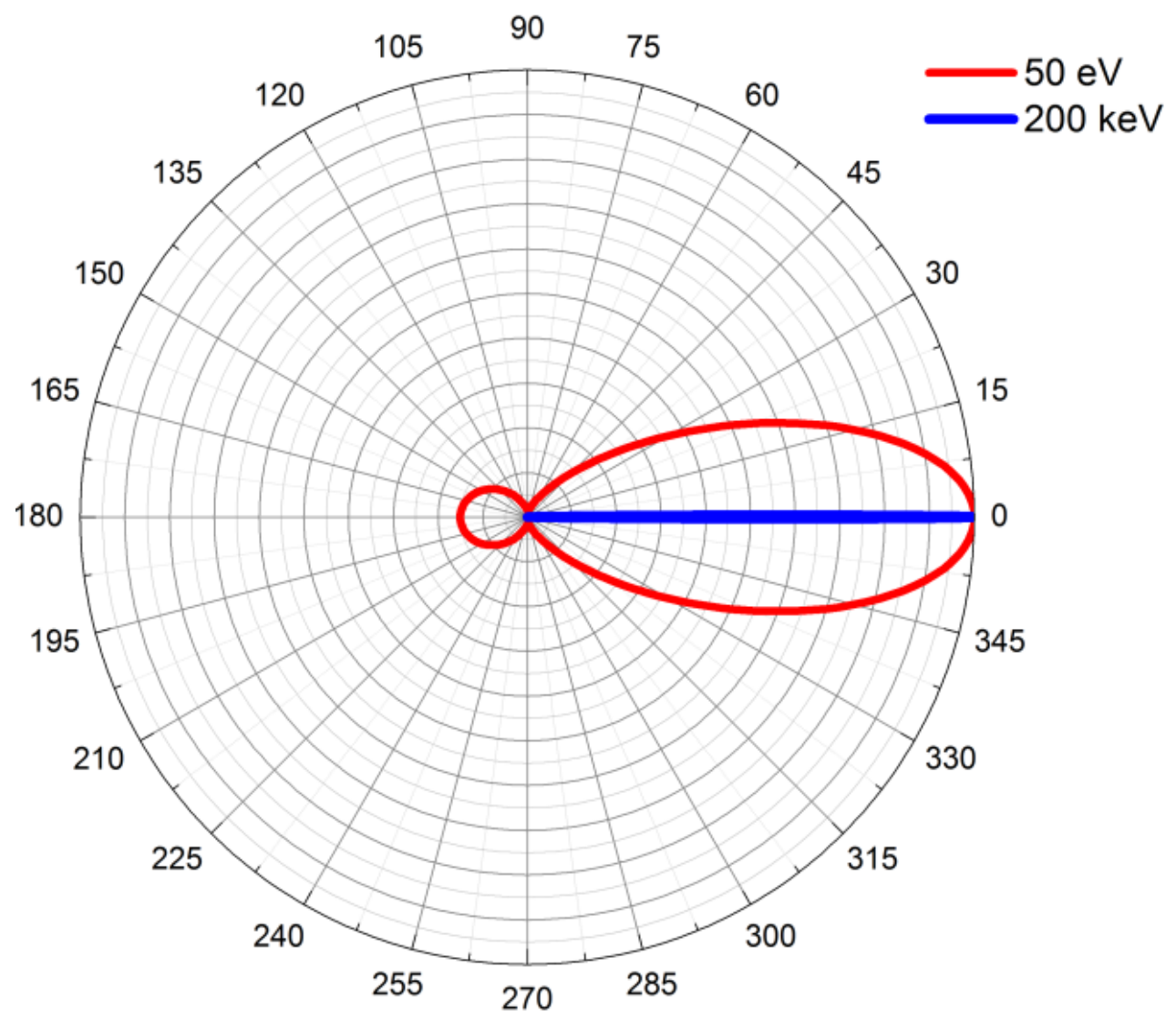

Fig. 3. Polar plot of the differential cross-section (DCS) distribution as a function of the scattering angle for electrons of 50 eV and 20 keV energies scattered at a carbon atom; calculated using the NIST database.

## 1.4 Large diffraction angle

Low-energy electrons diffract at larger angles than the electrons of conventional energies. The diffraction angle is approximately given by $\sin\vartheta \propto \lambda / d$, where $d$ is the characteristic size of the object. Figure 4 shows the distribution of the amplitude of an electron waves of 300 keV and 100 eV diffracted from a point-like object. For 300 keV energy electrons, the diffracted wave remains "focused" even at a distance of tens of nanometers, while for 100 eV energy electrons, the diffracted wave is already "defocused" at a distance less than one nanometer. This has consequences for imaging of three-dimensionally distributed objects like proteins. When a low-energy electron wave propagates through a protein, at the exit plane, one observes a "defocused" image of the protein, with a strong contrast in the amplitude and phase distributions [2], Fig. 5. For 300 keV electrons, on the other hand, the exit wave shows almost no contrast in the amplitude distribution, but contrast in the phase distribution. Since only intensity (amplitude) can be recorded, the macromolecules are conventionally imaged in defocus, like, for example, it is done in cryo-electron microscopy.

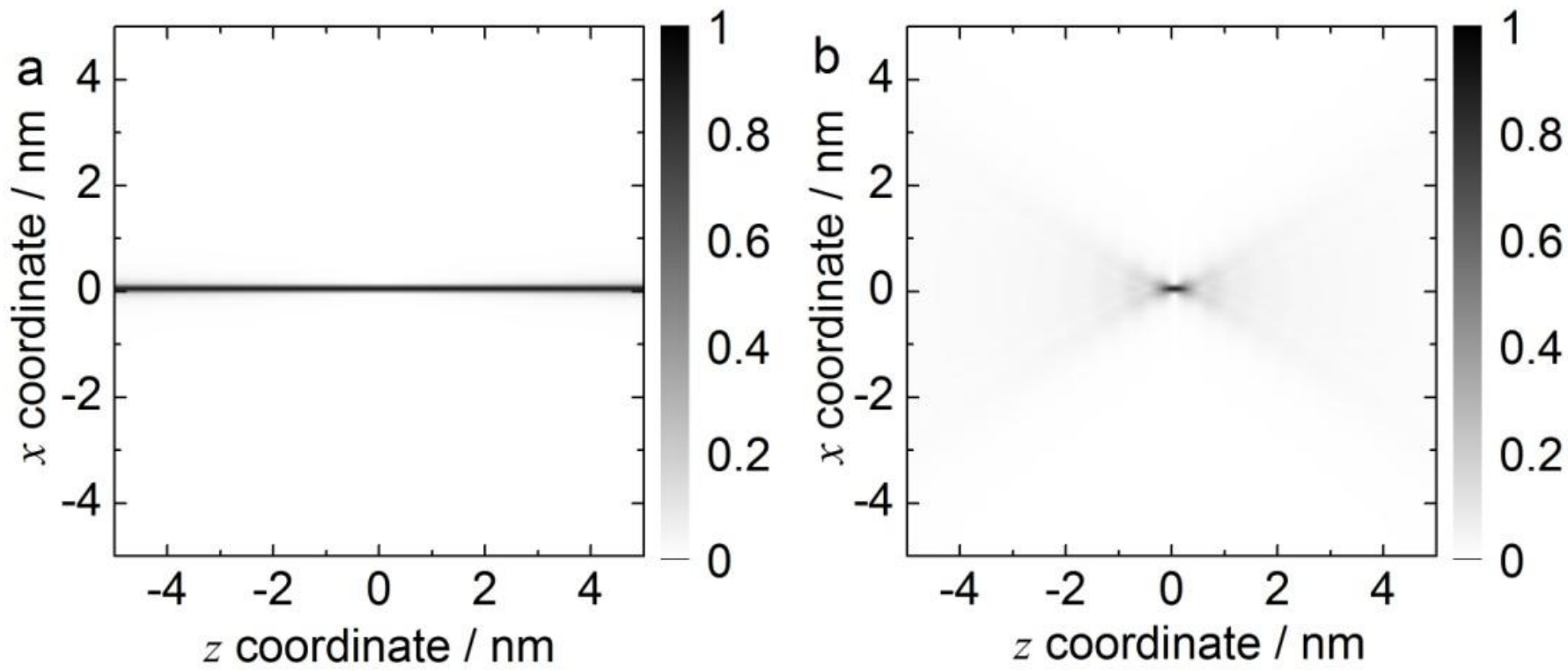

Fig. 4. Amplitude distributions (contrast inverted) of the point spread functions (PSF) for (a) 300 keV and (b) 100 eV energy electrons scattered by a point-like object. Adapted from [2].

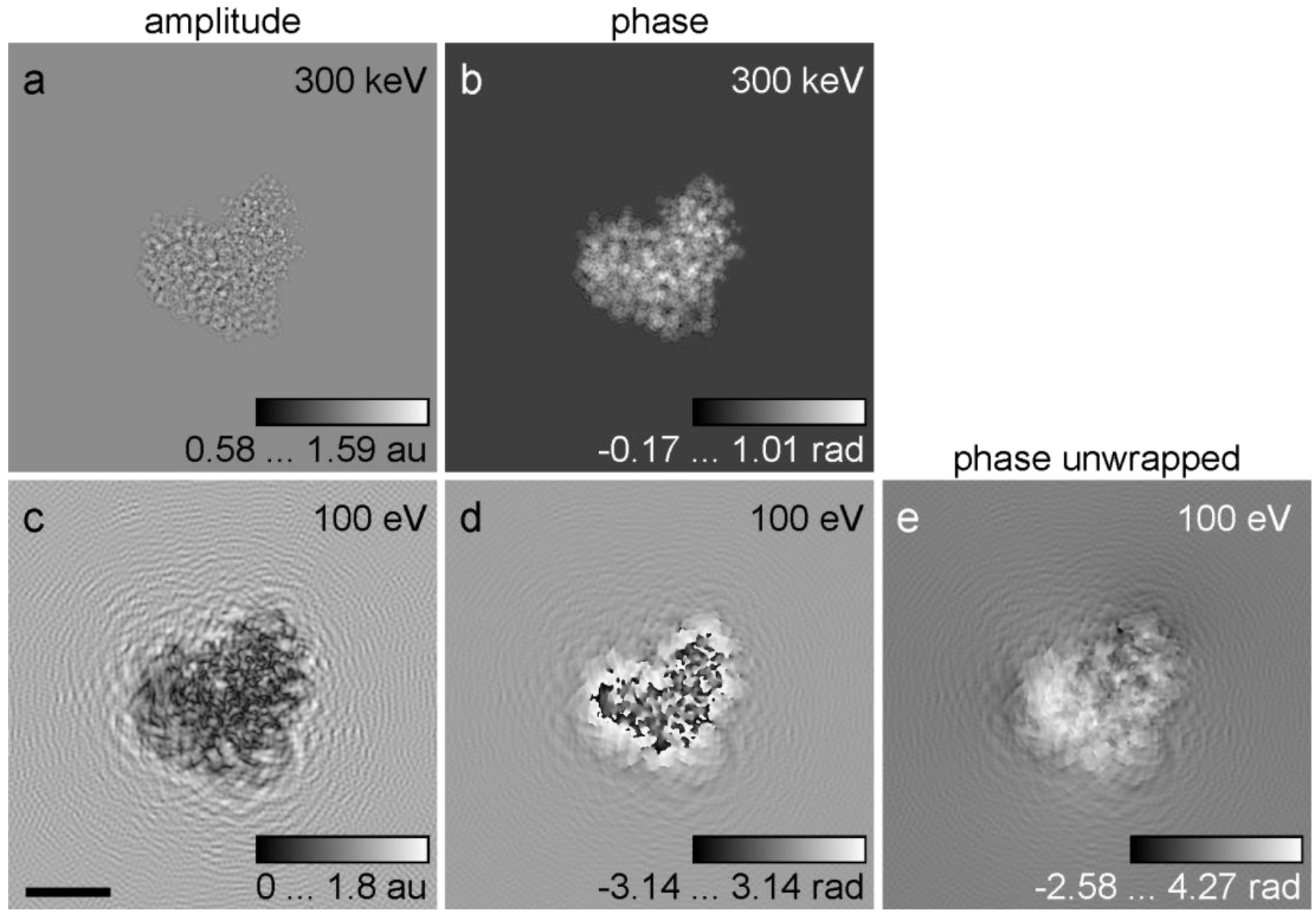

Fig. 5. Simulated complex-valued distribution of (a) – (b) 300 keV and (c) – (e) 100 eV electron exit wave after propagating through a lysozyme molecule. Scalebar, 1 nm. Adapted from [2].

## 1.5 Low radiation damage to biological samples

Ultra low-energy electrons have been reported to cause very low radiation damage to biological specimens, which makes them suitable for imaging single biological macromolecules such as proteins. It was demonstrated that a bundle of free-standing DNA molecules was continuously exposed to low-energy electrons for 70 minutes without any detected radiation damage, Fig. 6. In the experiment, 20 holograms were recorded every 10 minutes. The cross-correlation coefficient, calculated (in a selected) region between the first 20 holograms and each hologram set during the 70 minutes, remained nearly one, which means that there was no detectable radiation damage [9]. If single DNA molecules were put into a conventional TEM, they would be immediately destroyed. Various individual biological macromolecules were imaged using low-energy electron holography: single DNA molecules [9-11], purple membrane [7, 12], tobacco mosaic virus [13, 14], bacteriophage [15], and others [16-18].

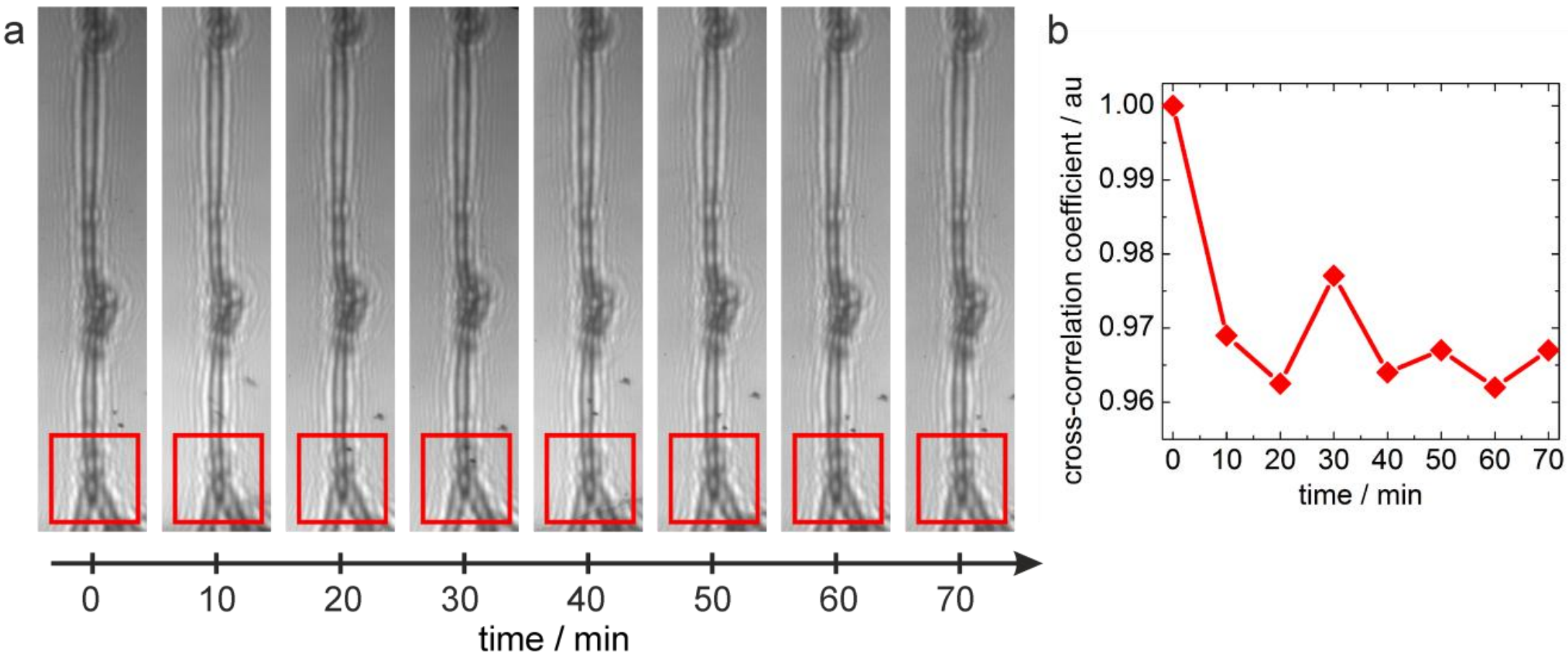

Fig. 6. Non-destructive imaging of DNA. The low-energy electron hologram of DNA molecules stretched over a hole in a thin film imaged continuously for 70 minutes using

electrons of 60 eV kinetic energy and a total current of 200 nA is shown in (a). A region in the holograms, marked in red, has been chosen to evaluate the cross-correlation function of subsequent holograms. The evolution of the cross-correlation coefficient is shown in (b).

## 1.6 High sensitivity to local charges and potentials

Low-energy electrons are more sensitive to the local charges and potentials than electrons of conventional energies. An adsorbate with an elementary charge sitting on top of graphene causes about 20% of contrast change when imaged with low-energy electrons, while there is less than 1% of contrast change when imaged by electrons with conventional energies, Fig. 7 [19, 20].

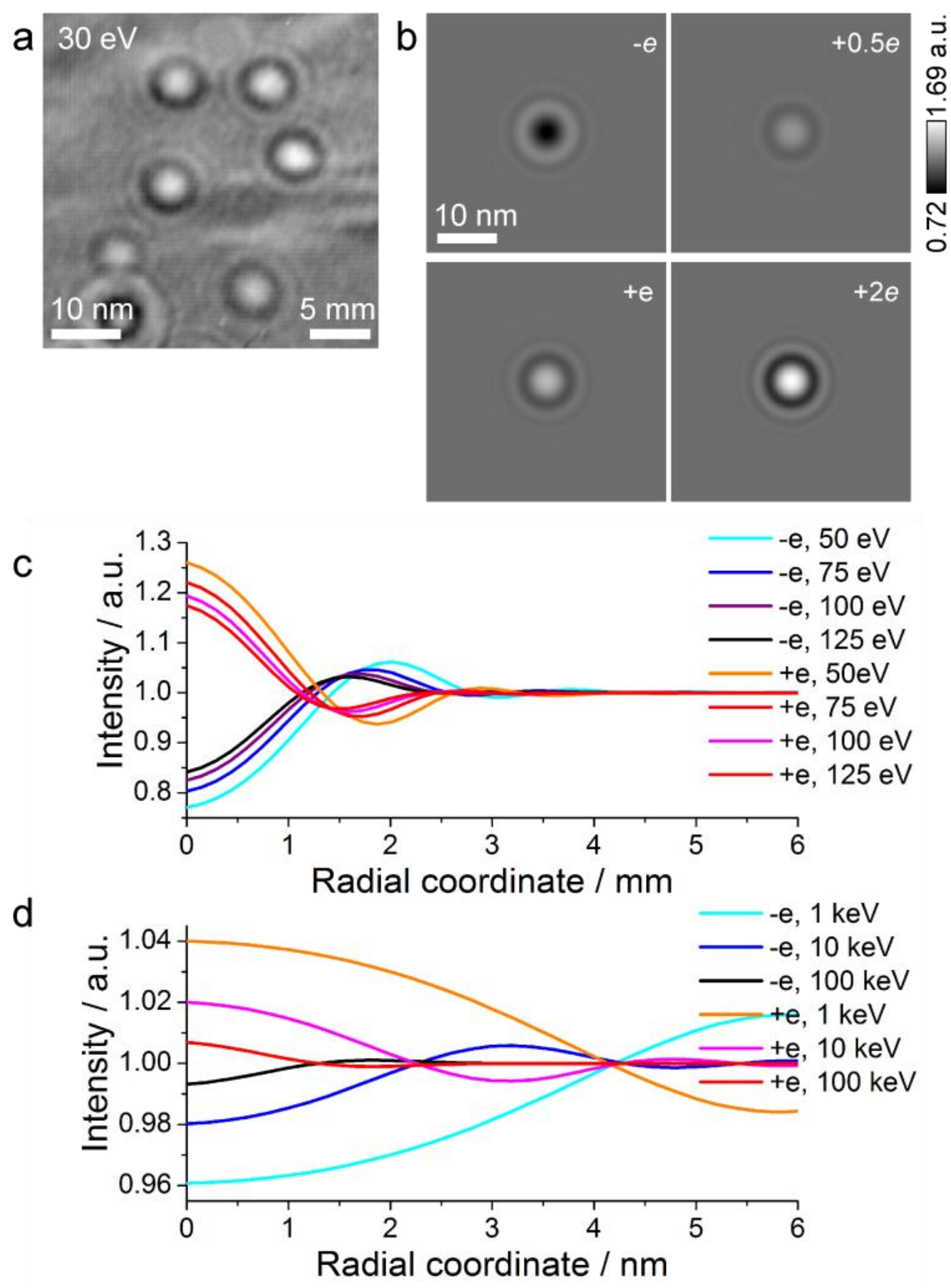

Fig. 7. Low-energy electron holograms of charged adsorbates. (a) A hologram exhibiting bright spots, recorded with 30 eV electrons, at a source-to-detector distance of 47 mm and a source-to-sample distance of 82 nm, at a resolution of 0.6 nm. (b) Simulated holograms of a point charge of four different charge values at an electron energy of 30 eV at a source-to-detector distance of 47 mm and a source-to-sample distance of 82 nm. (c) and (d) the angular averaged intensity profiles as a function of the radial coordinate of the holograms of point changes imaged with low- and high-energy electrons.

This high sensitivity to local charges and potentials can be exploited for imaging charges in biological macromolecules. Holograms of single DNA molecules exhibit regions with extremely bright and dark intensity. These regions are due to some charging in the DNA: a positive charge deflects the electrons inwards, thus acting like a small lens, while a negative charge deflects electrons outwards, creating a region with no signal (dark region) [11].

### 1.7 Strong phase shift

Low-energy electrons are slower than electrons of conventional energies. As a result, they spend much longer time in the sample potential and thus acquire a larger phase shift than electrons of conventional energies; this effect can be seen in the phase distributions shown in Fig. 5. However, objects with strong phase shift are known to be more difficult for reconstruction from their diffraction patterns or holograms [21].

### 1.8 Sumup of the properties of ultra low-energy electrons

The properties of low-energy electrons are summarized in Table 1.

| advantages | disadvantages |
|---|---|
| low radiation damage to biological samples | short IMFP (~ 5 Å) |
| imaging of individual macromolecules | very thin samples |
| high sensitivity to local potentials | large phase shift – difficult for phase retrieval |

Table 1. Properties of low-energy electrons.

## 2. Experimental imaging techniques that use low-energy electrons

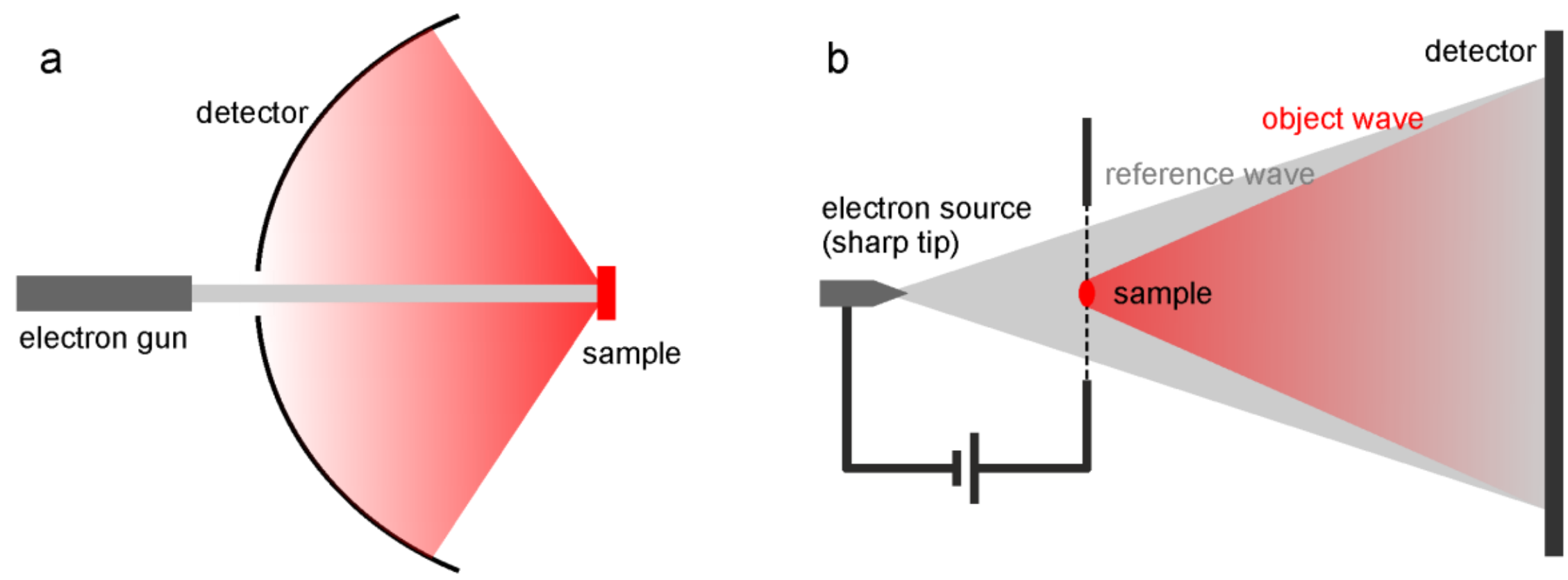

Fig. 5. Schemes of experimental imaging techniques that use low-energy electrons. (a) Low-energy electron diffraction (LEED). (b) Low-energy electron holography (LEEH).

### 2.1 Low-energy electron diffraction (LEED)

The combination of short IMFP and strong backward scattering amplitude makes low-energy electrons a perfect radiation to probe surfaces in backscattering mode. In LEED, an electron gun shoots electrons onto the surface of the sample, and backscattered electrons are collected by a spherical detector positioned behind the source, Fig. 5a. LEED is perfect for surface characterization, because only a few surface layers are probed.

### 2.2 Low-energy electron holography (LEEH)

LEEH is realized in transmission mode by using the holography principle. Historically, holography was proposed by Gabor in 1948 (patented in 1947) [22, 23]. At that time, the first electron microscopes became available, but there was a problem: despite the short wavelength, one did not observe atomic resolution in the obtained images. The reason was outlined by Scherzer – the unavoidable aberrations due to the lenses [24]. Gabor came up with an ingenious idea – instead of trying the performance of the lenses, one can just remove all the lenses between the object and the detector. In this case, one part of the electron wave is scattered by the object (object wave) $O$, and the other part is not scattered by the object (reference wave) $R$, Fig. 5b. The two waves interfere in the detector plane, and the resulting interference pattern (hologram) given by:

$$H = |O + R| \propto OR^* + O^* R \quad (5)$$

contains the entire information (phase and amplitude) of the object wave without any aberrations. For his invention of holography, Gabor was awarded a Nobel Prize in 1971, after lasers were invented and holography became a practical, popular toy.

In our research, we develop low-energy electron holography and use a custom-built low-energy electron microscope [25], sketched in Fig. 5b. The electron source is a sharp tungsten tip; the sharper the tip, the better is the coherence of the emitted electron wave. The electrons are extracted by field emission. The source-to-sample distance can be varied between 20 and 1000 nanometers, which, in turn, varies the energy of the electrons. The magnification of the microscope is about $10^5$, allowing imaging of nano-scaled objects. The theoretical resolution calculated as $R = \lambda / 2NA$, where $NA$ is the numerical aperture of the microscope, gives 2.5 Ångstrom at the energy 150 eV, source-to-detector distance of 18 cm and detector diameter of 75 mm.

The samples are numerically reconstructed from the recorded holograms the same way as in optical holography. Firstly, the hologram is illuminated by the reference wave. From Eq. (5), it gives the object wave in the hologram plane:

$$RH \propto O|R|^2 . \quad (6)$$

Then, the obtained wavefront $O$ is propagated backward from the detector plane to the sample plane. The distance of the propagation can be selected, which enables reconstruction of the sample distribution at different planes and quasi three-dimensional reconstruction [26].

Low-energy electrons in a LEEH microscope can be collimated into a parallel beam by using a micro-lens [27, 28], which enables diffraction imaging. In this arrangement, diffraction patterns of individual TMV macromolecules [29] and twisted bilayer graphene where satellite peaks due to moiré structure are observed [30] have been measured.

The current resolution of LEEH is about 7 – 8 Å when imaging biological macromolecules [17]. It was also demonstrated that single atoms on top of graphene can be resolved by LEEH [31]. LEEH can be realized in live mode – charge redistribution in single DNA molecules [11] and in situ intercalation of alkali atoms in bilayer graphene [32] were observed by LEEH. At very short source-to-sample distances, the band structure of graphene can be directly visualized by LEEH [33].